\documentclass[envcountsect]{llncs}
\usepackage[pdftex]{graphicx}
\graphicspath{{figures/}}
\DeclareGraphicsExtensions{.jpg,.png}

\usepackage[caption=false,font=footnotesize]{subfig}
\usepackage{floatrow}
\usepackage[font=footnotesize,labelfont=bf]{caption}

\usepackage{multirow}
\usepackage{makecell}

\usepackage{amsmath}
\usepackage{bbm}
\usepackage{stmaryrd}

\usepackage{array}
\usepackage{color}
\usepackage[hyphens]{url}
\usepackage[pdftitle=Title,pdfauthor=Anonymous]{hyperref}
\usepackage{adjustbox}
\usepackage{booktabs}
\usepackage{multirow}
\usepackage{rotating}
\usepackage{makecell}
\usepackage{hhline}
\usepackage{lscape}
\usepackage{tabu}
\usepackage{tikz}
\usepackage{textcomp} 
\usetikzlibrary{shapes,backgrounds} 

\usepackage{colortbl}

\usepackage{listings, xcolor}

\definecolor{verylightgray}{rgb}{.97,.97,.97}
\lstdefinelanguage{Solidity}{
keywords=[1]{anonymous, assembly, assert, balance, break, call, callcode, case, catch, class, constant, continue, contract, debugger, default, delegatecall, delete, do, else, event, export, external, false, finally, for, function, gas, if, implements, import, in, indexed, instanceof, interface, internal, is, length, library, log0, log1, log2, log3, log4, memory, modifier, new, payable, pragma, private, protected, public, pure, push, require, return, returns, revert, selfdestruct, send, storage, struct, suicide, super, switch, then, this, throw, transfer, true, try, typeof, using, value, view, while, with, addmod, ecrecover, keccak256, mulmod, ripemd160, sha256, sha3}, 
	keywordstyle=[1]\color{blue}\bfseries,
	keywords=[2]{Stages,States, address, bool, byte, bytes, bytes1, bytes2, bytes3, bytes4, bytes5, bytes6, bytes7, bytes8, bytes9, bytes10, bytes11, bytes12, bytes13, bytes14, bytes15, bytes16, bytes17, bytes18, bytes19, bytes20, bytes21, bytes22, bytes23, bytes24, bytes25, bytes26, bytes27, bytes28, bytes29, bytes30, bytes31, bytes32, enum, int, int8, int16, int24, int32, int40, int48, int56, int64, int72, int80, int88, int96, int104, int112, int120, int128, int136, int144, int152, int160, int168, int176, int184, int192, int200, int208, int216, int224, int232, int240, int248, int256, mapping, string, uint, uint8, uint16, uint24, uint32, uint40, uint48, uint56, uint64, uint72, uint80, uint88, uint96, uint104, uint112, uint120, uint128, uint136, uint144, uint152, uint160, uint168, uint176, uint184, uint192, uint200, uint208, uint216, uint224, uint232, uint240, uint248, uint256, var, void, ether, finney, szabo, wei, days, hours, minutes, seconds, weeks, years},	
	keywordstyle=[2]\color{teal}\bfseries,
	keywords=[3]{block, blockhash, coinbase, difficulty, gaslimit, number, timestamp, msg, data, gas, sender, sig, value, now, tx, gasprice, origin},	
	keywordstyle=[3]\color{violet}\bfseries,
	identifierstyle=\color{black},
	sensitive=false,
	comment=[l]{//},
	morecomment=[s]{/*}{*/},
	commentstyle=\color{gray}\ttfamily,
	stringstyle=\color{red}\ttfamily,
	morestring=[b]',
	morestring=[b]"
}

\usepackage{xspace}

\newcommand{\etc}{\textit{etc.}\xspace}
\newcommand{\ie}{\textit{i.e.,}\xspace}
\newcommand{\eg}{\textit{e.g.,}\xspace}

\newcommand{\cmmnt}[1]{\ignorespaces} 

\begin{document}
\frontmatter
\mainmatter

\title{SoK: Transparent Dishonesty: Front-running Attacks on Blockchain.}

\author{
	Shayan Eskandari\inst{\dag\ddag},
	Seyedehmahsa Moosavi\inst{\dag},
	Jeremy Clark\inst{\dag}
	}

\institute{
	\textsuperscript{\dag} Gina Cody School of Engineering and Computer Science\\ Concordia University \\
	\textsuperscript{\ddag} ConsenSys Diligence
	}

\maketitle



\begin{abstract}

We consider \textit{front-running} to be a course of action where an entity benefits from prior access to privileged market information about upcoming transactions and trades. Front-running has been an issue in financial instrument markets since the 1970s. With the advent of the blockchain technology, front-running has resurfaced in new forms we explore here, instigated by blockchain's decentralized and transparent nature.
In this paper, we draw from a scattered body of knowledge and instances of front-running across the top 25 most active decentral applications (DApps) deployed on Ethereum blockchain. Additionally, we carry out a detailed analysis of \textsf{Status.im} initial coin offering (ICO) and show evidence of abnormal miner's behavior indicative of front-running token purchases. Finally, we map the proposed solutions to front-running into useful categories. 
\end{abstract}





\section{Introduction} \label{sec:intro}

Blockchain technology enables decentralized applications (DApps) or smart contracts. Function calls (or transactions) to the DApp are processed by a decentralized network. Transactions are finalized in stages: they (generally) first relay around the network, then are selected by a miner and put into a valid block, and finally, the block is well-enough incorporated that is unlikely to be reorganized. Front-running is an attack where a malicious node observes a transaction after it is broadcast but before it is finalized, and attempts to have its own transaction confirmed before or instead of the observed transaction.

The mechanics of front-running work on all DApps but front-running is not necessarily beneficial, depending on the DApp's internal logic and/or as any mitigations it might implement. Therefore, DApps need to be studied individually or in categories. In this paper, we draw from a scattered body of knowledge regarding front-running attacks on blockchain applications and the proposed solutions, with a series of case studies of DApps deployed on Ethereum (a popular blockchain supporting DApps). We do case studies on decentralized exchanges (\eg Bancor), crypto-collectibles (\eg CryptoKitties), gambling services (\eg Fomo3D), and decentralized name services (\eg Ethereum Name Service). We also study initial coin offerings (ICOs). Finally, we provide a categorization of techniques to eliminate or mitigate front-running including transaction sequencing, cryptographic techniques like commit/reveal, and redesigning the functioning of the DApp to provide the same utility while removing time dependencies.




\section{Preliminaries \& Related Work}

\subsection{Traditional Front-running}
\label{sec:What is front-running?}

\emph{Front-running} is a course of action where someone benefits from early access to market information about upcoming transactions and trades, typically because of a privileged position along the transmission of this information and is applicable to both financial and non-financial systems. Historically, floor traders might have overheard a broker's negotiation with her client over a large purchase, and literally race the broker to buy first, potentially profiting when the large purchase temporarily reduces the supply of the stock. Alternatively, a malicious broker might front-run their own client's orders by purchasing stock for themselves between receiving the instruction to purchase from the client and actually executing the purchase (similar techniques can be used for large sell orders). Front-running is illegal in jurisdictions with established securities regulation.

Cases of front-running are sometimes difficult to distinguish from related concepts like insider trading and arbitrage. In front-running, a person sees a concrete transaction that is set to execute and reacts to it before it actually gets executed. If the person instead has access to more general privileged information that might predict future transactions but is not reacting at the actual pending trades, we would classify this activity as insider trading. If the person reacts after the trade is executed, or information is made public, and profits from being the fastest to react, this is considered arbitrage and is legal and encouraged because it helps markets integrate new information into prices quickly.

\subsection{Literature on Traditional Front-running}\label{traditionalFrontrunning}
Front-running originates on the Chicago Board Options Exchange (\textit{CBoE}) \cite{markham1988front}. The Securities Exchange Commission \textit{(SEC)} in 1977 defined it as: ``The practice of effecting an options transaction based upon non-public information regarding an impending block transaction\footnote{A block in the stock market is a large number of shares, 10\,000 or more, to sell which will heavily change the price.} in the underlying stock, in order to obtain a profit when the options market adjusts to the price at which the block trades.~\cite{sec1978optionsmarket}'' Self-regulating exchanges (\eg \textit{CBoE}) and the \textit{SEC} spent the ensuing years planning how to detect and outlaw front-running practices~\cite{markham1988front}. The \textit{SEC} stated: ``It seems evident that such behaviour on the part of persons with knowledge of imminent transactions which will likely affect the price of the derivative security constitutes an unfair use of such knowledge.\footnote{Securities Exchange Act Release No. 14156, November 19, 1977, (Letter from George A. Fitzsimmons, Secretary, Securities, and Exchange Commission to Joseph W. Sullivan, President  CBoE).}'' The \textit{CBoE} tried to educate their members on existing rules, however, differences in opinion regarding the unfairness of front-running activities, insufficient exchange rules and lack of a precise definition in this area resulted in no action~\cite{sec1978optionsmarket} until the SEC began the regulation. We refer the reader interested in further details on this early regulatory history to Markham~\cite{markham1988front}. The first front-running policies applied only to certain option markets. In 2002, the rule was expanded to cover all security futures~\cite{finra_2002}. In 2012, it was expanded further with the new amendment, FINRA Rule 5270, to cover trading in options, derivatives, or other financial instruments overlying a security with only a few exceptions~\cite{sec2012frontrunning,finra_2012}. Similar issues have been seen with domain names~\cite{sac022en33:online,edelman2009front} as well.


\subsection{Background on Blockchain Front-running} \label{sec:Front Running on the Blockchains}

%
\begin{figure}[t]
\floatbox[{\capbeside\thisfloatsetup{capbesideposition={left,top},capbesidewidth=0.3\linewidth}}]{figure}[\FBwidth]
{\caption{The front-runner upon spotting the profitable transaction \textit{Buy(1000)} sends his own transaction with higher gas price to bribe the miners to prioritize his transaction over initial transaction.}\label{fig:RegularFrontrunning}}
{\includegraphics[width=1.0\linewidth]{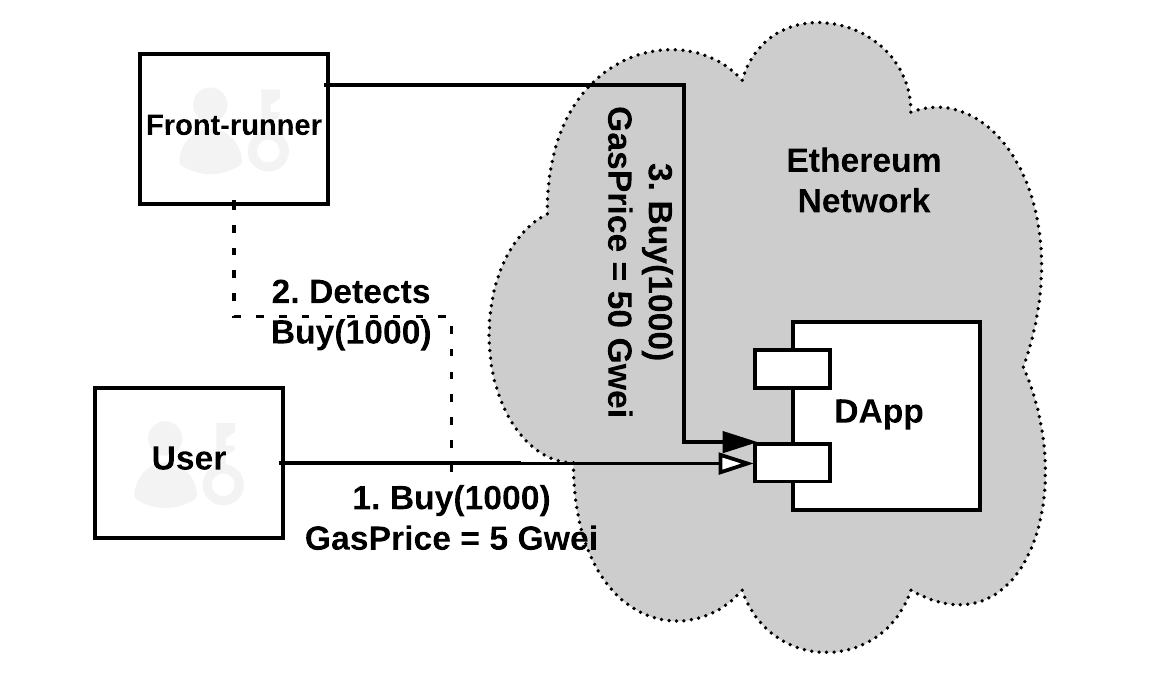}}
\end{figure}

Blockchain technology (introduced via Bitcoin in 2008~\cite{nakamoto2008bitcoin}) strives to disintermediate  central parties that participate in a transaction. However, blockchains also introduce new participants in the process of relaying and finalizing transactions. Miners are in the best position to conduct these attacks as they hold fine-grained control over the exact set of transactions that will execute and in what order and can mix in their own (late) transactions without broadcasting them. Miners do however have to commit to what their own transactions will be before beginning the proof of work required to solve a block.

Any user monitoring network transactions (\eg running a full node) can see unconfirmed transactions. On the Ethereum blockchain, users have to pay for the computations in a small amount of Ether called \textsf{gas}~\cite{AccountT67:online}. The price that users pay for transactions, \textsf{gasPrice}, can increase or decrease how quickly miners will execute them and include them within the blocks they mine. A profit-motivated miner who sees identical transactions with different transaction fees will prioritize the transaction that pays a higher gas price due to limited space in the blocks. This has been called a gas auction~\cite{frontrunme}. Therefore, any regular user who runs a full-node Ethereum client can front-run pending transactions by sending adaptive transactions with a higher gas price (see Figure~\ref{fig:RegularFrontrunning}).

Finally, well-positioned relaying nodes on the network (or part of the broader internet backbone) can attempt to influence how transactions are propagated through the network, which can influence the order miners receive transactions, or if they receive them at all~\cite{HKZG15,MGH18}.

\subsection{Literature on Blockchain Front-running}

Given the purpose of this entire paper is systemizing the existing literature, we do not re-enumerate the literature here. However, we note two points. First, we are not aware of any other systematic study of this issue. Second, front-running is related to two well-studied concepts: double-spending and rushing adversaries~\cite{kosba2016hawk}.




Double-spending attacks in Bitcoin are related to front-running~\cite{bamert2013have,karame2012double}. In this attack, a user broadcasts a transaction and is able to obtain some off-blockchain good or service before the transaction has actually been (fully) confirmed. The user can then broadcast a competing transaction that sends the same unspent coins to herself, perhaps using higher transaction fees, arrangements with miners or artifacts of the network topology to have the second transaction confirmed instead of the first. This can be considered a form of \textsf{self-front-running}.
In the cryptographic literature, front-running attacks are modeled by allowing a so called `rushing' adversary to interact with the protocol~\cite{beaver1992cryptographic}. In particular, ideal functionalities of blockchains (such as those used in simulation-based proofs) need to capture this adversarial capability, assuming the real blockchain does not address front-running. See \eg Bitcoin backbone~\cite{garay2015bitcoin} and Hawk~\cite{kosba2016hawk}.


\section{A Taxonomy of Front-running Attacks}
\label{sec:taxonomy}

As we will illustrate with examples through-out the paper, front-running attacks can often be reduced to one of a few basic templates. We emphasize what the adversary is trying to accomplish (without worrying about how) and we distinguish three cases: displacement, insertion, and suppression attacks. In all three cases, Alice is trying to invoke a function on a contract that is in a particular state, and Mallory will try to invoke her own function call on the same contract in the same state before Alice.

In the first type of attack, a \textit{displacement attack}, it is not important to the adversary for Alice's function call to run after Mallory runs her function. Alice's can be orphaned or run with no meaningful effect. Examples of displacement include: Alice trying to register a domain name and Mallory registering it first~\cite{kalodner2015empirical}; Alice trying to submit a bug to receive a bounty and Mallory stealing it and submitting it first~\cite{breidenbach2018enter}; and Alice trying to submit a bid in an auction and Mallory copying it. 

In an \textit{insertion attack}, after Mallory runs her function, the state of the contract is changed and she needs Alice's original function to run on this modified state. For example, if Alice places a purchase order on a blockchain asset at a higher price than the best offer, Mallory will insert two transactions: she will purchase at the best offer price and then offer the same asset for sale at Alice's slightly higher purchase price. If Alice's transaction is then run after, Mallory will profit on the price difference without having to hold the asset.

In a \textit{suppression attack}, after Mallory runs her function, she tries to delay Alice from running her function. After the delay, she is indifferent to whether Alice's function runs or not. We only observe this attack pattern in one DApp and the details are quite specific to it, so we defer discussion until Section~\ref{sec:gambling}.

Each of these attacks have two variants, \textit{asymmetric} and \textit{bulk}. In some cases, Alice and Mallory are performing different operations. For example, Alice is trying to cancel an offer, and Mallory is trying to fulfill it first. We call this \textit{asymmetric displacement}. In other cases, Mallory is trying to run a large set of functions: for example Alice and others are trying to buy a limited set of shares offered by a firm on a blockchain. We call this \textit{bulk displacement}. 


\section{Cases of Front-running in DApps}



\definecolor{UnitedNationBlue}{rgb}{0.30,0.53,1}
\definecolor{LightSteelBlue}{rgb}{0.69,0.77,0.87}
\definecolor{LightGrey}{rgb}{0.83,0.83,0.83}

\begin{table}[t]
\captionsetup{justification = centering, singlelinecheck = false}
{\caption{Top 25 DApps based on recent user activity from \texttt{DAppRadar.com} on September 4th, 2018. The DApps that are in bold are discussed in this paper.}\label{tab:top25DApps}}
\begin{tabular}{|c|c|c|l|}

\hline
\rowcolor{LightSteelBlue}
 \textbf{DApp Category}    										& \textbf{Names} & Rank \\  \hline
\multirow{6}{*}{Exchanges} 										& IDEX & 1 \\ 
															& \cellcolor{LightGrey} \textbf{ForkDelta, EtherDelta} & 2 \\
															& \cellcolor{LightGrey} \textbf{Bancor} & 7 \\
															& The Token Store & 13 \\
															& LocalEthereum & 14 \\
															& Kyber & 22 \\ 
															& \cellcolor{LightGrey} \textbf{0x Protocol} & 23 \\ \hline

\multirow{9}{*}{\shortstack{Crypto-Collectible \\ Games \\ (ERC-721~\cite{erc721})}}
															& \cellcolor{LightGrey} \textbf{CryptoKitties} & 3 \\  
															& Ethermon & 4\\
															& Cryptogirl & 9\\
															& Gods Unchained TCG & 12\\
															& Blockchain Cuties & 15\\
															& ETH.TOWN! & 16\\
															& 0xUniverse & 18\\
															& MLBCrypto Baseball & 19\\
															& HyperDragons & 25\\ \hline

\multirow{8}{*}{Gambling}										& \cellcolor{LightGrey} \textbf{Fomo3D} & 5 \\  
															& DailyDivs & 6 \\													& PoWH 3D & 8	 \\ 
															& FomoWar & 10 \\
															& FairDapp & 11\\
															& Zethr & 17 \\
															& dice2.win & 20 \\ 
															& Ether Shrimp Farm & 21 \\  \hline

\multirow{1}{*}{Name Services}									& \cellcolor{LightGrey} \textbf{Ethereum Name Service} & 24  \\  \hline

\end{tabular}
\vspace{1em}
\end{table}


To find example DApps to study, we used the top 25 DApps based on recent user activity from \texttt{DAppradar.com} in September 2018.\footnote{List of decentralized applications \url{https://DAppradar.com/DApps}} User activity is admittedly an imperfect metric for finding the `most significant' DApps: significant DApps might be lower volume overall or for extended periods of time (\eg ICOs, which we remedy by studying independently in Section~\ref{sec:ICOsfrontrunning}). However, user activity is an objective criteria, data on it is available, and the list captures our intuition about which DApps are significant. It suffices for a first study in this area, and is preferable over an ad hoc approach. Using the dataset, we categorized the top 25 applications into 4 principal use cases. The details are given in Table~\ref{tab:top25DApps}.


\subsection{Markets and Exchanges} \label{sec:frontrunningExchanges}

\begin{figure}[t]
\floatbox[{\capbeside\thisfloatsetup{capbesideposition={right,bottom},capbesidewidth=0.3\linewidth}}]{figure}[\FBwidth]
{\caption{The adversarial miner monitors the Ethereum mempool for decentralized exchange transactions. Upon spotting a profitable cancellation transaction, he puts his buy order prior to the cancel transaction in the block he mines. Doing so, the miner can profit from the underlying trade and also get the gas included in the cancel transaction.}\label{fig:MinerFrontrunning}}
{\includegraphics[width=\linewidth]{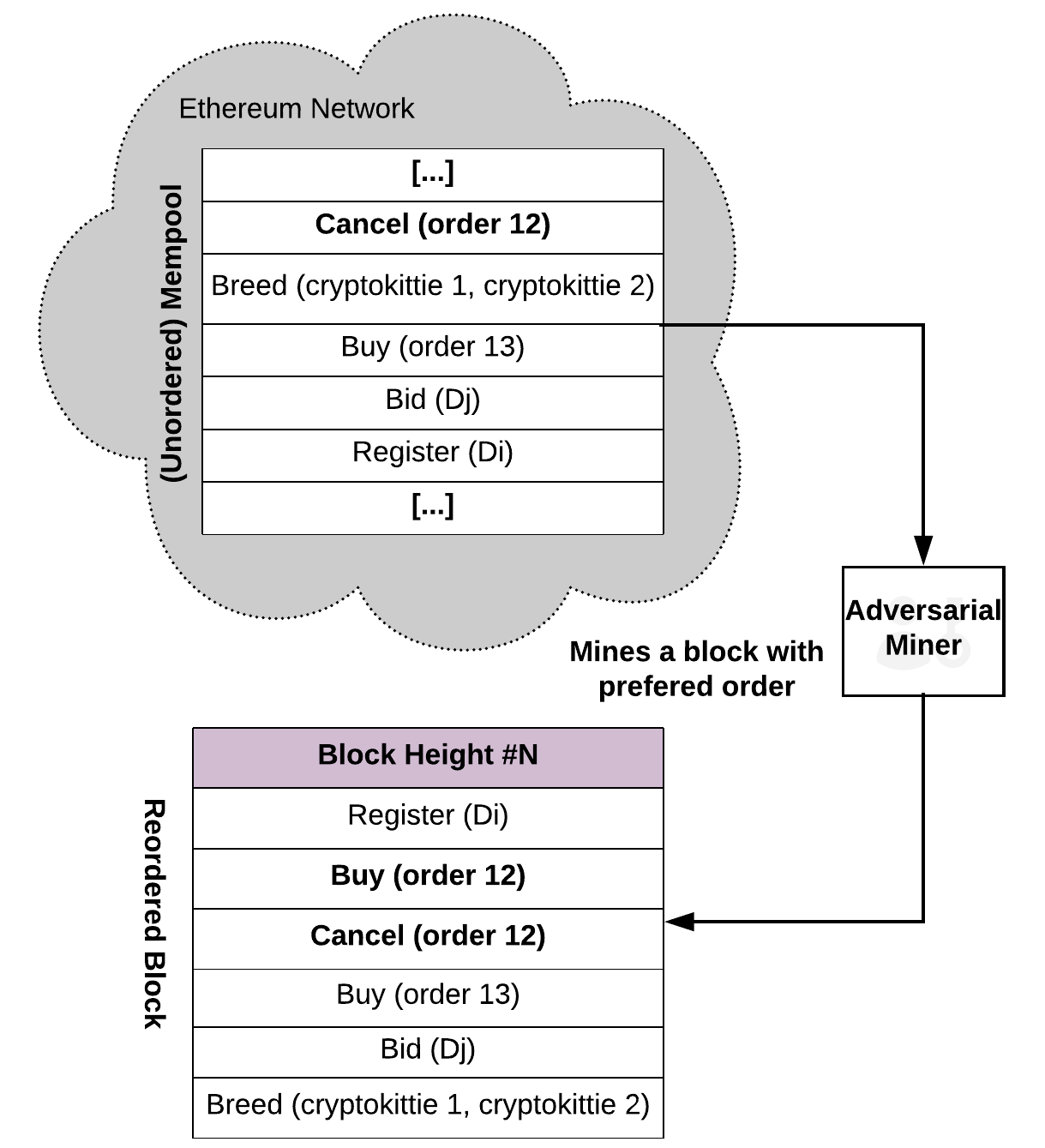}}
\end{figure}


The first category of DApp in Table~\ref{tab:top25DApps} are financial exchanges for trading ether and Ethereum-based tokens. Exchanges such as EtherDelta\footnote{Also known as ForkDelta for the user interface: \url{https://forkdelta.app/}}, purport to implement a decentralized exchange, however, their order books are stored on a central server they control and shown to their users with a website interface. Central exchanges can front-run orders in the traditional sense, as well as re-order or block orders on their servers. 0xProtocol~\cite{warren20170x} uses \textit{Relayers} which act as the order book holders and could front-run the orders they relay.

As seen in traditional financial markets, one method to manipulate the spot price of an asset, is to flood the market with orders and cancel them when there are filling orders (``taker's griefing''~\cite{consesnsys0xReview2017}). Placing an order in a partially centralized exchange is free, but to prevent taker's griefing attacks, the user needs to send an Ethereum transaction to cancel each of his orders. Cancelling orders is most important when prices change faster than order execution. In this case, when an adversarial actor sees a pending cancellation transaction, he sends a fill order transaction with higher gasPrice to get in front of the cancellation order and take the order before it is canceled (this is known as \textit{cancellation grief}). This attack follows the asymmetric displacement template and is illustrated in Figure~\ref{fig:MinerFrontrunning}. 

Designing truly decentralized exchanges, where the order book is implemented directly on a public blockchain, is being pursued by a number of projects~\cite{inDEXgithub}. These designs are generally vulnerable to front-running attacks following a displacement or insertion template. For example, a front-running full node or miner might gauge the demand for trades at a given price by the number of pending orders, and try to displace them at the same price assuming the demand is the result of the accurate new information about the asset. Alternatively,  the front-runner might observe a large market order (\ie it will execute at any price). The adversary can try to insert a pair of limit orders that will bid near the best offer price and offer at a higher price. If the pair executes ahead of the market order, the front-runner profits by scalping the price of the shares. Finally, if adversary has pre-existing offers likely to be reached by the market order, she could insert cancelations and new offers at a higher price.


Bancor is an exchange DApp that allows users to exchange their tokens without any counter-party risk. The protocol aims to solve the cryptocurrency liquidity issue by introducing \textit{Smart Tokens}~\cite{hertzog2017bancor}. Smart tokens are ERC20-compatible that can be bought or sold through a DApp-based dealer that is always available and implements a market scoring rule to manage its prices. Bancor provides continuous liquidity for digital assets without relying on brokers to match buyers with sellers. Implemented on the Ethereum blockchain, when transactions are broadcast to the network, they sit in a pending transaction pool known as \textit{mempool} waiting for the miners to mine them. Since Bancor handles all the trades and exchanges on the chain (unlike other existing decentralized exchanges), these transactions are all visible to the public for some time before being included within a block. This leaves Bancor vulnerable to the blockchain race condition attack as attackers are given enough time to front-run other transactions, in which they can gain favourable profits by buying before the order and fill the original order with slightly higher price~\cite{BancorIs7:online}. Researchers have shown and implemented a proof of concept code to front-run Bancor as a non-miner user~\cite{NewTab13:online}.


\subsection{Crypto-Collectibles Games}\label{sec:cryptogames}

The second category of DApp in Table~\ref{tab:top25DApps} is crypto-collectables. Consider Cryptokitties~\cite{cryptokitties}, the most active DApp in this category and third most active overall. Each kitty is a cartoon kitten with a set of unique features to distinguish it from other cryptokitties, some features are rarer and harder to obtain. They can be bought, sold, or bred with other cryptokitties. At the Ethereum level, the kitty is a token implemented with \textit{ERC-721: Non-Fungible Token Standard}~\cite{erc721}. Kitties are generally bought and sold on-chain through auction smart contracts. See Sections ~\ref{sec:frontrunningExchanges} and ~\ref{sec:ens} for more details on auction-based front-running attacks.


Specific to Cryptokitties protocol, they can breed and give birth. When cryptokitties breed, the smart contract sets from which future block the pregnancy of the cat can be completed. Anyone can complete the pregnancy by calling \texttt{giveBirth()} after the birthing block and they will receive a reward in ether\footnote{As there are no automated function calls in Ethereum, this incentive model --known as \textit{Action Callback}~\cite{klerosGeneralizedFrontrunner2019}-- is used to encourage users to call these functions.}. Even though front-running these calls would not affect the protocol workflow, but this displacement attack could result in financial profit for front-runners~\cite{zhou2018erays,cryptoMidwivesBot2018}.


\subsection{Gambling}
\label{sec:gambling}

The third category of DApp in Table~\ref{tab:top25DApps} is gambling services. While a large category of gambling games are based on random outcomes, DApps do not have unique access to an unpredictable data stream to harvest for randomness~\cite{pierrot2018malleability}. Any candidate source of randomness (such as block headers) is accessible to all DApp functions and can also be manipulated to an extent by miners.

\textsf{Fomo3D} is an example of a game style (known as \textsf{Exit Scam}\footnote{\url{https://exitscam.me/play}}) not based on random outcomes, and it is the most active game on Ethereum in our sample. The aim of this game is to be the last person to have purchased a ticket when a timer goes to zero in a scenario where anyone can buy a ticket and each purchase increases the timer by 30 seconds. Many speculated such a game would never end but on August 22, 2018, the first round of the game ended with the winner collecting 10,469 Ether\footnote{The first winner of Fomo3D, won 10,469 Ether \url{ https://etherscan.io/tx/0xe08a519c03cb0aed0e04b33104112d65fa1d3a48cd3aeab65f047b2abce9d508}} equivalent to \$2.1M USD at the time.
Blockchain forensics indicate a sophisticated winning strategy to displace any new ticket purchases~\cite{fomo3dhacker,fomo3dmedium} that would reset the counter. The winner appears to have started by deploying many high gas consumption DApps unrelated to the game. When the timer of the game reached about 3 minutes, the winner bought 1 ticket and then sent multiple high gasPrice transactions to her own DApps. These transactions congested the network and bribed miners to prioritize them ahead of any new ticket purchases in \textsf{Fomo3D}. Recall this basic form of bribery is called a \textit{Gas Auction}; See related work ~\cite{mccorry2018smart,bonneau2016buy} for more sophisticated bribery contracts. 

We classify this in the unique category of a suppression attack in our taxonomy (see Section~\ref{sec:taxonomy}). At first glance, it seemed like an extreme version of an asymmetric/bulk displacement attack on any new ticket purchase transactions. However the key difference is that the front-runner does not care at all about the execution of her transactions---if miners mined empty blocks for three minutes, that would also be acceptable. Thus, bulk displacement\footnote{Also known as Block Stuffing Attack~\cite{blockstuffing18}} is simply a means-to-an-end and not the actual end goal of the adversary.


\subsection{Name Services}\label{sec:ens}

The final category in Table~\ref{tab:top25DApps} is name services, which are primarily aimed at disintermediating central parties involved in web domain registration (\eg ICAAN and registrars) and resolution (\eg DNS). For simple name services (such as some academic work like Ghazal~\cite{moosavi2018ghazal}), domains purchases are transactions and front-runners can displace other users attempting to register domains. This parallels front-running attacks seen in regular (non-blockchain) domain registration~\cite{sac022en33:online}. \textsf{Ethereum Name Service (ENS)}~\cite{ensEIP} is the most active naming service on Ethereum. Instead of allowing new \texttt{.eth} domain names to be purchased directly, they are put up for a sealed bid auction which seals the domain name in a bid, but not the bid amount. Most implementations use the more user friendly but less confidential method for starting and bidding on a domain name: \texttt{startAuctionsAndBid()}. This method leaks the hash of the domain and the initial bid amount in the auction. Original names can be guessed from the hashes (\eg rainbow tables, used in ENS Twitter bot\footnote{\url{https://twitter.com/ensbot}}) or people can bid on domains even though they do not know what they are because of speculation on its value. 

Users are allowed to bid for 3 days before the 2-day reveal phase begins (see ~\ref{sec:comm}), in which all bidders (winners and losers) must send a transaction to reveal their bids for a specific domain or sacrifice their bid amount . Also note that if two bidders bid the same price, the first to reveal wins it~\cite{ENSHandlingFrontRunningDiscussion}. Using the leaked information, the domain squatter can win the auction with the same price of the original bidder by revealing it first. This is similar to front-running as it relies on inserting an action before the user, however we do not consider this specific action as front-running attack.


\section{Cases of Front-running in ICOs}  \label{sec:ICOsfrontrunning}

Initial coin offerings (ICOs) have changed how blockchain firms raise capital. More than 3000 ICOs have been held on Ethereum, and the market capitalization of these tokens appears to exceed \$75B USD in the first half of 2018~\cite{zetzsche2018ico}. At the DApp level, tokens are offered in short-term sales that see high transaction activity while the sale is on-going and then the activity tapers off to occasional owner transfers. When we collected the top 25 most active DApps on \texttt{DAppRadar.com}, no significant ICOs were being sold. The ICO category slips through our sampling method, but we identify it as a major category of DApp and study it here.

\subsection{\textit{Status.im} ICO}
To deal with demand, ICOs cap sales in a variety of ways to mitigate front-running attacks. In June 2017, \textit{Status.im}~\cite{statuswhitepaper} started its ICO and reached the predefined cap within 16 hours, collecting close to 300,000 Ether. In order to prevent wealthy investors purchasing all the tokens and limit the amount of Ether deposited in each investment, they used a \textit{fair} token distribution method called \textit{Dynamic Ceiling} as an attempt to increase the opportunity for smaller investors. They implemented multiple caps (ceilings) in which, each had a maximum amount that could be deposited in. In this case, every deposit was checked by the smart contract and the exceeding amount was refunded to the sender while the accepted amount was sent to their multi-signature wallet address~\cite{statusicoanalysis}.

During the time frame the ICO was open for participation, there were reports of Ethereum network being unusable and transactions were not confirming. Further study showed that some mining pools might have been manipulating the network for their own profit. In addition, there were many transactions sent with a higher gas price to front-run other transactions, however, these transactions were failing due to the restriction in the ICO smart contract to reject transactions with higher than 50 \textit{GWei} gas price (as a mitigation against front-running).

\begin{figure}[t]
\floatbox[{\capbeside\thisfloatsetup{capbesideposition={left,bottom},capbesidewidth=0.35\linewidth}}]{figure}[\FBwidth]
{\caption{The percentage of Ethereum blocks mined between block 3903900 and 3908029, this is the time frame in which Status.im ICO was running. This percentage roughly shows the hashing power ratio each miner had at that time.}\label{fig:mining_pool_ratio}}
{\includegraphics[width=1.1\linewidth]{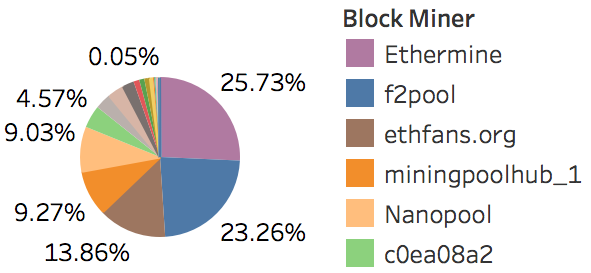}}
\end{figure}

\subsection{Data Collection and Analysis}
According to the analysis we carried out, we discovered that the F2Pool---an Ethereum mining pool that had around 23\% of the mining hash rate at the time (Figure~\ref{fig:mining_pool_ratio})---sent 100 Ether to 30 new Ethereum addresses before the Status.im ICO started. When the ICO opened, F2Pool constructed 31 transactions to the ICO smart contract from their addresses, without broadcasting the transactions to the network\footnote{Note that we do not have an authoritative copy of the mempool over time, however, the probability of these transactions being broadcasted to the network and exclusively get mined by the same pool as the sender is low.}. They used their entire mining power to mine their own transactions and some other potentially failing high gas price transactions.

Ethereum's blockchain contains all transaction ever made on Ethereum. While the default client and online blockchain explorers offer some limited query capabilities, in order to analyze this case, we built our own database. Specifically, we used open source projects such as Go Ethereum implementation\footnote{Official Go implementation\url{https://github.com/ethereum/go-ethereum}.} for the full node, a python script for extracting, transforming and loading Ethereum blocks, named \texttt{ethereum-etl}~\cite{ethereumetl} and Google BigQuery.\footnote{\url{https://cloud.google.com/bigquery/}} Using this software stack, we were able to isolate transactions within the Status.im ICO. We used data analysis tool \texttt{Tableau}.\footnote{\url{https://www.tableau.com/}} A copy of this dataset and the initial findings can be found in our Github repository\footnote{\url{http://bit.ly/madibaFrontrunning}}.

\begin{figure}[t]
\floatbox[{\capbeside\thisfloatsetup{capbesideposition={left,top},capbesidewidth=0.3\linewidth}}]{figure}[\FBwidth]
{\caption{This chart shows the miners behaviour on the time frame that Status.im ICO was running. It is clear that the number of successful transactions mined by F2Pool do not follow the random homogeneous pattern of the rest of the network.}\label{fig:Transactions_miners_while_status_ico_cut}}
{\includegraphics[width=1.1\linewidth]{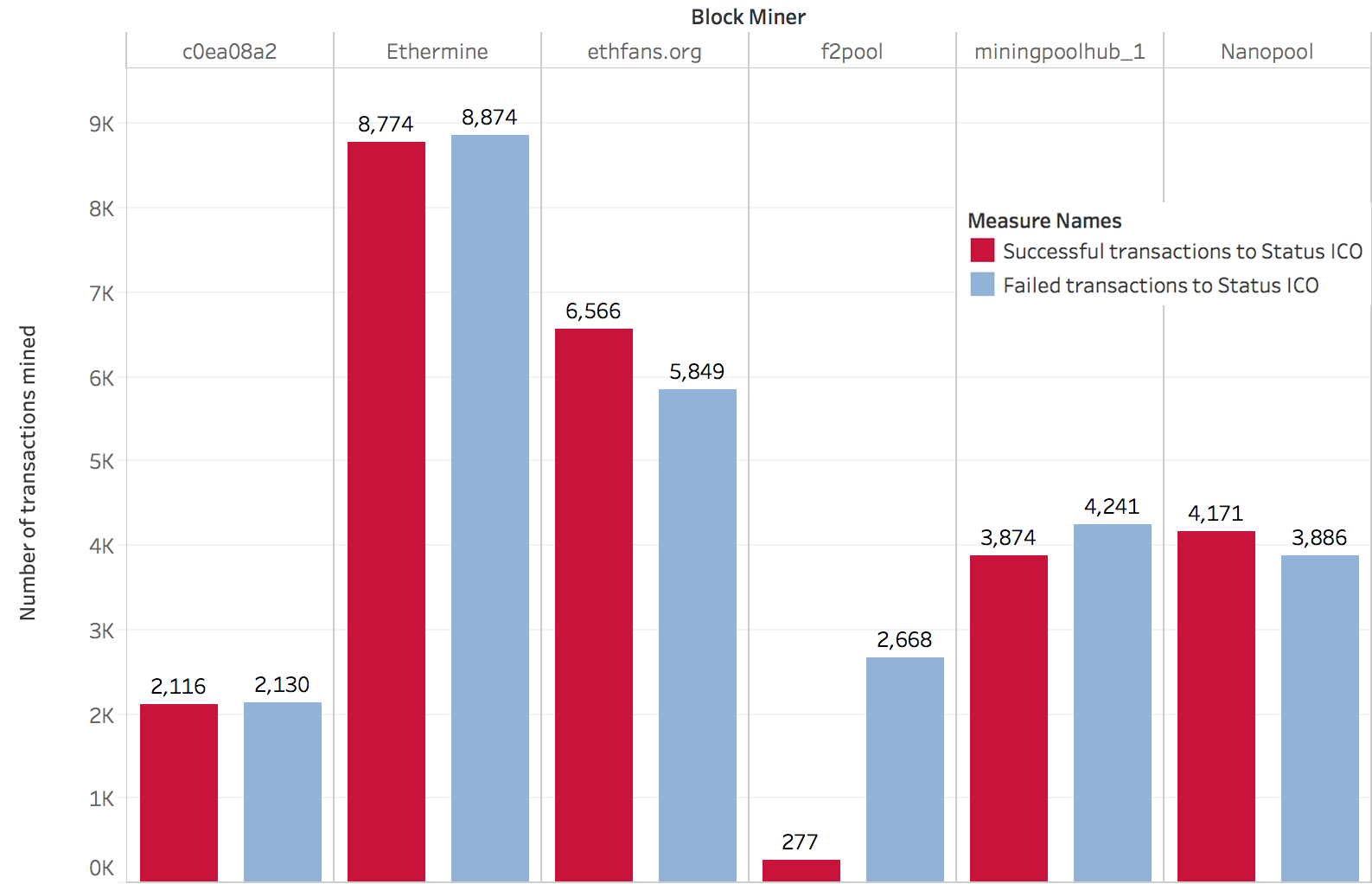}}
\end{figure}

As shown in Figure~\ref{fig:Transactions_miners_while_status_ico_cut}, most of the top miners in the mentioned time frame, have mined almost the same number of failed and successful transactions which were directed toward Status.im token sale, however F2Pool's transactions indicate their successful transactions were equivalent to 10\% of the failed transactions, hence maximizing the mining rewards on gas, while censoring other transactions to the token sale smart contract. The terminology used here is specific to smart contract transactions on Ethereum, by \textit{``failed transaction''} we mean the transactions in which the smart contract code rejected and threw an exception and by \textit{``successful transaction''} we mean the transactions that went through and received tokens from the smart contract.

\begin{figure}[t]
\floatbox[{\capbeside\thisfloatsetup{capbesideposition={right,bottom},capbesidewidth=0.5\linewidth}}]{figure}[\FBwidth]
{\caption{Prior to \textit{Status.im} ICO  \textit{F2Pool} deposited 100 Ether in multiple new Ethereum addresses. On the time of the ICO, transactions sent from these addresses to \textit{Status} ICO smart contract were prioritized in their mining pool, resulting in purchasing \textit{ERC20} tokens. This method was used to overcome the dynamic ceiling algorithm of the ICO smart contract. Later on they sent the refunded Ether back to their own address.\protect\footnotemark \label{fig:f2poolfront-run}}}
{\includegraphics[width=0.8\linewidth]{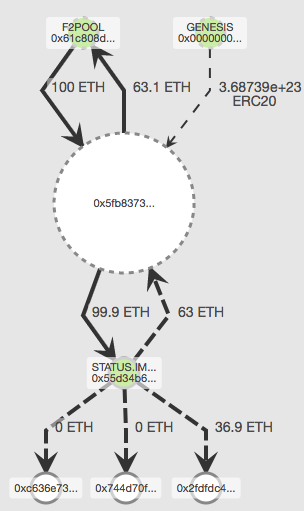}}
\end{figure}

\footnotetext{Graph was made using \url{Blockseer.com} blockchain explorer.}

By tracing the transactions from these 30 addresses, we found explicit interference by F2Pool\footnote{F2Pool address was identified by their mining reward deposit address \url{https://etherscan.io/address/0x61c808d82a3ac53231750dadc13c777b59310bd9}.} in this scenario. As shown in Figure~\ref{fig:f2poolfront-run}, the funds deposited by F2Pool in these addresses were sent to \textit{Status.im} ICO and mined by F2Pool themselves, where the dynamic ceiling algorithm refunded a portion of the deposited funds. A few days after these funds were sent back to F2Pool main address and the tokens were aggregated later in one single address. Although this incident does not involve transaction reordering in the blocks, it shows how miners can modify their mining software to behave in a certain way to front-run other transactions by \textit{bulk displacement} to gain monetary profit.


\section{Key Mitigations} 


As we studied front-running attacks on the blockchain, we also encountered a number of ways of preventing, detecting or mitigating front-running attacks. Instead of providing the details of exact solutions which will change over time, we extract the main principles or primitives that address the attack. A particular system may implement more than one in a layered mitigation approach.

We classify the mitigations into three main categories. In the first category, the blockchain removes the miner's ability to arbitrarily order transactions and tries to enforce some ordering, or queue, for the transactions. In the second category, cryptographic techniques are used to limit the visibility of transactions, giving the potential front-running less information to base their strategy on. In the final category, DApps are designed from the bottom-up to remove the importance of transaction ordering or time in their operations. We also note that for DApps that are legally well-formed (\eg with identified parties and a clear jurisdiction), front-running attacks can violate laws, which is its own deterrent. 

\paragraph{Traditional Front-running Prevention Methods.} There are debates in traditional markets regarding the fact that front-running is considered to be a form of insider trading which deemed to be illegal. Traditional methods to prevent front-running mainly involves after the fact investigation and legal action against the front-runners~\cite{FTFrontrunning18}. As mentioned in section~\ref{traditionalFrontrunning}, defining front-running and educating the employees were the first step taken to prevent such issues in traditional markets, however, front-running became less likely to happen mainly because of the high fine and lawsuits against firms who behaved in an unethical way. Other methods such as dark pools~\cite{zhu2014dark,buti2011diving} and sealed bids~\cite{radner1989sealed} were discussed and implemented in a variety of regulated trading systems. The traditional methods to prevent front-running does not apply to blockchain applications, as mainly they are based on central enforcement and limitations, also in case of blockchains the actors who are front-running could be anonymous and the fear of lawsuits would not apply. 

\subsection{Transaction Sequencing}

Ethereum miners store pending transactions in pools and draw from them when forming blocks. As the term `pool' implies, there is no intrinsic order to how transactions are drawn and miners are free to sequence them arbitrarily.\footnote{Sometimes the pool is called a `queue.' It is important to note is a misnomer as queues enforce a first-in-first-out sequence.} The vanilla Go-Ethereum (geth) implementation prioritizes transactions based on their gas price and nonce~\cite{EthMinerTxOrder}. Because no rule is enforced, miners can sequence transactions in advantageous ways. A number of proposals attempt to thwart this attack by enforcing a rule about how to sequence transactions.

First-in-first-out (FIFO) is generally not possible on a distributed network because transactions can reach different nodes in a different order. While the network could theoretically form a consensus based on locally observed FIFO, this would increase the rate of orphaned blocks, as well as adding complexity to the protocol. A trusted third party can be used to assign sequential numbers to transactions (and sign them), but this is contrary to blockchain's core innovation of distributed trust. Nonetheless, some exchanges do centralize time-sensitive functionalities (\eg  \textit{EtherDelta} and \textit{0xProject}) in off-chain order books~\cite{warren20170x,0xfrontrunning:online}.

One alternative is to sequence transactions pseudorandomly. This can be seen in proposals like Canonical Transaction Ordering Rule (CTOR) by Bitcoin Cash ABC~\cite{bitcoincashabcFork} which adds transactions in lexicographical order according to their hash~\cite{bitcoinABC2018CTOR}. Note that Bitcoin does not have a front-running problem for standard transactions. While this could be used by Ethereum to make front-running statistically difficult, the protection is marginal at best and might even exacerbate attacks. A front-runner can construct multiple equivalent transactions, with slightly different values, until she finds a candidate that positions her transaction a desirable location in the resulting sequence. She broadcasts only this transaction and now miners that include her transaction will position it in front of transactions they heard about much earlier.

Finally, transactions themselves could enforce order. For example, they could specify the current state of the contract as the only state to execute on. This transaction chaining only prevents certain types of front-running; \ie it prevents insertion attacks but not displacement attacks (recall our taxonomy in Section~\ref{sec:taxonomy}). As transaction chaining only allows one state-changing transaction per state, at most one of a set of concurrent transactions can be confirmed; a drawback for active DApps.


\subsection{Confidentiality}

\paragraph{Privacy-Preserving Blockchains.}

All transaction details in Bitcoin are made public and participant identities are only lightly protected. A number of techniques increase confidentiality~\cite{bunzBulletproofs,maxwell2015confidential} and anonymity ~\cite{miers2013zerocoin,cryptoeprint2015,sasson2014zerocash} for cryptocurrencies. A current research direction is extending these protections to DApps~\cite{aztec2018,ZoeEth2017}. It is tempting to think that a confidential DApp would not permit front-running, as the front-runner would not know the details of the transaction she is front-running. However, there are some nuances here to explore.

A DApp interaction includes the following components: (1) the code of the DApp, (2) the current state of the DApp, (3) the name of the function being invoked, (4) the parameters supplied to the function, (5) the address of the contract the function is being invoked on, and (6) the identity of the sender. Confidentiality applied to a DApp could mean different levels of protection for each of these. For front-running, function calls (3,4) are the most important, however, function calls could be inferred from state changes (2). Hawk~\cite{kosba2016hawk} and Ekiden~\cite{cheng2018ekiden} are examples of (2,3,4)-confidentiality (with limitations we are glossing over).

The applicability of privacy-preserving blockchains needs to be evaluated on a case-by-case basis. For example, one method used by traditional financial exchanges in dealing with front-running from high frequency traders is a dark pool: essentially a (2,3,4)-confidential order book maintained by a trusted party. A DApp could disintermediate this trusted party. Users whose balances are affected by changes in the contract's state would need to be able to learn this information. Further, if the contract addresses are known (\ie no 5-confidentiality), front-runners can know about the traffic pattern of calls to contracts which could be sufficient grounds for attack; for example, if each asset on an exchange has its own market contract, this leaks trade volume information. As a contrasting example, consider again decentralized domain registration: hiding state changes (2-confidentiality) defeats the entire purpose of the DApp, and protecting function calls is ineffective with a public state change since the state itself reveals the domain being registered.

\paragraph{Commit/Reveal.}
\label{sec:comm}

While confidentiality appears insufficient for solving domain name front-running alone, a hybrid approach of sequencing and confidentiality can be effective and is, in fact, an example of an older cryptographic trick known as commit/reveal. The essence of the approach is to protect the function call (\eg (3,4)- or (4)-confidentiality) until the function is enqueued in a sequence of functions to be executed. Once the sequence is established, the confidentiality is lifted and the function can only be executed in the order it was enqueued (or, generally speaking, not at all).

\begin{figure}[t]
\floatbox[{\capbeside\thisfloatsetup{capbesideposition={left,bottom},capbesidewidth=0.5\linewidth}}]{figure}[\FBwidth]
{\caption{Commit and Reveal. User sends a commitment transaction with the hash of the data, After the commitment period is over, user sends her reveal transaction to the DApp revealing the information that matches the commitment.}\label{fig:commitReveal}}
{\includegraphics[width=\linewidth]{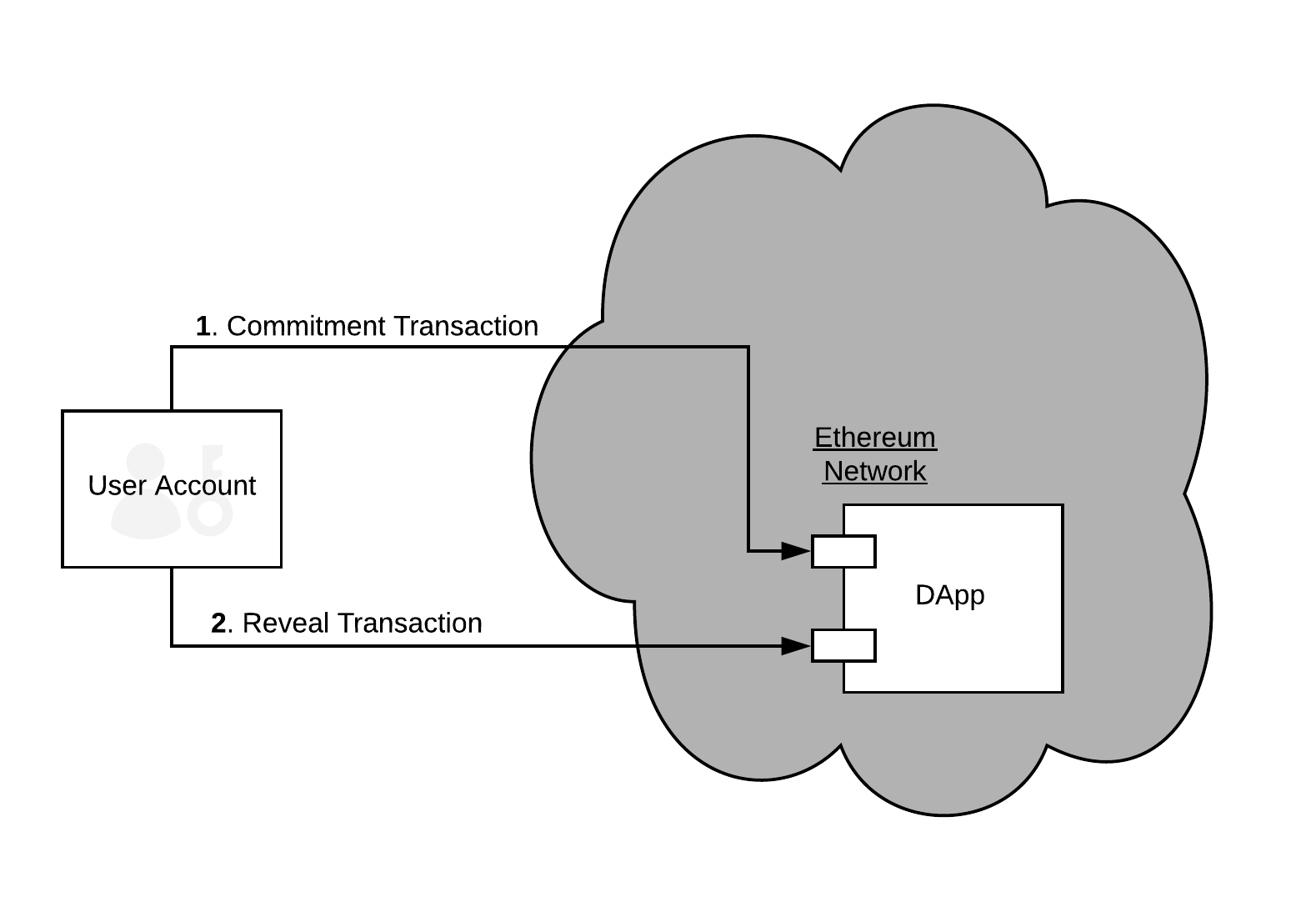}}
\end{figure}

Recall that a commitment scheme enables one to commit to a digital value (\eg a statement, transaction, data, \etc) while keeping it a secret (\textit{hiding}), and then open it (and only it: \textit{binding}) at a later time of the committer's choosing~\cite{brassard1988minimum}. A common approach (conjectured to be hiding) is to submit the cryptographic hash of the value with a random nonce (for low entropy data) to a smart contract, and later reveal the original value and nonce which can be verified by the contract to correctly hash to the commitment (see Figure~\ref{fig:commitReveal}).

An early application of this scheme to blockchain is Namecoin, a Bitcoin-forked DApp for name services~\cite{kalodner2015empirical}. In Namecoin, a user sends a commit transaction which registers a new hidden domain name, similar to a sealed bid. Once this first transaction is confirmed, a time delay begins. After the delay, a second transaction reveals the details of the requested domain. This prevents front-running if the reveal transaction is confirmed faster than an adversarial node or miner can redo the entire process.

Commit/reveal is a two-round protocol, and aborting after the first round (early aborts) could be an issue for this (along with most multi-round cryptographic protocols). For example, in a financial exchange where the number of other orders might be in a predictable interval, an adversary can spray the sequence (\ie a price-time priority queue) with multiple committed transactions and no intention of executing them all. She then only reveal the ones that result in an advantageous trade.\footnote{This is analogous to behavior in traditional financial markets where high-frequency traders will make and cancel orders at many price points (flash orders or pinging). If they can cancel faster than someone can execute it---someone who has only seen the order and not the cancelation---then the victim reveals their price information.} There are other ways of aborting; if payments are required but not collateralized, the aborting party can ensure that payment is not available for transfer. One mitigation to early aborts that blockchain is uniquely positioned to make is having users post a fidelity bond of a certain amount of cryptocurrency that can be automatically dispensed if they fail to fully execute committed transactions (this is used in multi-round blockchain voting~\cite{mccorry2017smart}). Finally, we note that any multiple round protocol will have usability challenges: users must be aware that participating in the first round is not sufficient for completing their intention.

\begin{figure}[t]
\floatbox[{\capbeside\thisfloatsetup{capbesideposition={left,bottom},capbesidewidth=0.5\linewidth}}]{figure}[\FBwidth]
{\caption{Submarine Send~\cite{libsubmarine}. User generates an \textit{Unlock} transaction from which the commitment address is retrieved using ECDSA ECRecover. \textsf{1.} by funding the \textit{commitment address},user is committed to the transaction. \textsf{2.} User sends the \textit{reveal transaction} to the DApp, revealing the nature of the commitment transaction. \textsf{3.} She broadcasts the \textit{unlock transaction} to unlock the funds in the commitment address. \textsf{4.} After the \textit{"Auction''} is over, anyone can call \textit{Finalize} function to finalize the process.}\label{fig:LibSubmarine}}
{\includegraphics[width=\linewidth]{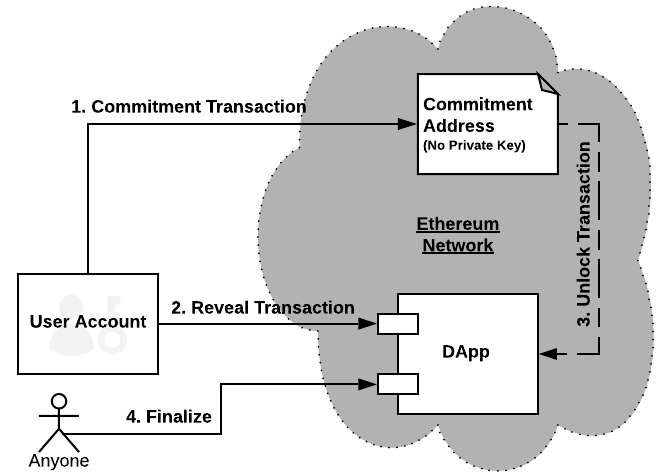}}
\end{figure}

\paragraph{Enhanced Commit/Reveal.} Submarine Commitments~\cite{libsubmarine,submarinesendHD} extend the confidentiality of the commit and reveal, so that the commitment transaction is identical to a transaction to a newly generated Ethereum address. They initially hide the contract address being invoked, providing (3,4,5)-confidentiality during the commit phase; and they ensure that if a revealed transaction sent funds, the funds were fully collateralized at commit time and are available to the receiving smart contract. See Figure~\ref{fig:LibSubmarine}.

\subsection{Design Practices}

The final main category of mitigation is to assume front-running is unpreventable and to thus responsively redesign the functionality of the DApp to remove any benefit from it. For example, when designing a decentralized exchange, one can use a call market design instead of a time-sensitive order book~\cite{clark2014decentralizing} to side-step and disincentivize front-running. In a call market design, the arrival time of orders does not matter as they are executed in batches\footnote{Also known as batch auctions~\cite{batchAuctions18}} . The call market solution pivots profitable gains that front-running miners stand to gain into fees that they collect~\cite{clark2014decentralizing}, removing the financial incentive to front-run.

In the finance literature, Malinova and Park discuss front-running mitigations for blockchain-based trading platforms~\cite{malinova2017market}. Instead of studying DApps, they develop an economic model where transactions, asset holdings, and traders' identities have greater transparency than in standard economic models---transparency they argue that could be accomplished by blockchain technology. However, in their model, they assume entities can interact directly over private channels to arrange trades. They define front-running in the context of private offers, where parties might adjust their position before accepting or countering a received offer. This model is quite different than the DApp-based model we study here. 

Another example in the design of ERC20 standard~\cite{erc20} is the allowance functionality. \textit{approve()} function in the specification allows a second entity to be able to spend N tokens from the sender's balance. In order to change the allowance, sender must send a transaction to set the new allowance value. Using the insertion attack, attacker could front-run the new allowance transaction and spend the old value before the new value is set~\cite{blockchainProjectAllowanceAttack,erc20GithubIssue}, and then additionally spend the new amount at a later time. Solutions such as \textit{decreaseApproval()}/\textit{increaseApproval()} were added in updated implementations.


\section{Concluding Remarks}

Front-running is a pervasive issue in Ethereum DApps. DApp developers don't necessarily have the mindset to design DApps with front-running in mind. This is an attempt to bring forward the subject and increase awareness of these type of attacks. While some DApp-level application logic could be built to mitigate these attacks, its ubiquity across different DApp categories suggests mitigations at the blockchain-level would perhaps be more effective. We highlight this as an important research area.



\subsubsection*{Acknowledgements.}
The authors thank the Autorit\'e des March\'es Financiers (AMF) for sponsoring this research through the Education and Good Governance Fund (EGGF), as well as NSERC through a Discovery Grant. 


\bibliography{bib/Frontrunning_Ref}



\end{document}